\documentclass[prb,twocolumn,aps,showpacs]{revtex4}
\usepackage{graphicx}
\def\ltsim{\vbox {\hbox{\lower .8\baselineskip \hbox{$<$}} \break
                 \hbox{\lower 0.2\baselineskip \hbox{$\sim$}} } }

\begin{document}

\title{Entanglement and Criticality in Quantum Impurity Systems}

\author{Karyn Le Hur,$^{1,2}$ Philippe Doucet-Beaupr\'e,$^2$ and Walter Hofstetter$^3$}
\affiliation{$^1$ Department of Physics, Yale University, New Haven, CT 06520, USA}
\affiliation{$^2$ D\'epartement de Physique, Universit\'e de Sherbrooke, 
Sherbrooke, Qu\'ebec, Canada J1K 2R1}
\affiliation{$^3$ Institut f\" ur Theoretische Physik, Johann Wolfgang Goethe-Universit\"at, 60438 Frankfurt/Main, Germany}
 
 \date{\today} 

\begin{abstract}
We investigate the entanglement between a spin and its environment in impurity systems which exhibit a second-order quantum phase transition. As an application, we employ the spin-boson model, describing a two-level system (spin) coupled to a subohmic bosonic bath with power-law spectral density, ${\cal J}(\omega)\propto \omega^s$ and $0<s<1$. Combining Wilson's Numerical Renormalization Group method and hyperscaling relations, we demonstrate that the entanglement between the spin and its environment is always enhanced at the quantum phase transition resulting in a visible cusp (maximum) in the entropy of entanglement. We formulate a  correspondence between criticality and impurity entanglement entropy, and the relevance of these ideas to Nano-systems is outlined.
\end{abstract}

\pacs{73.43.Nq, 72.15.Qm, 03.65.Ud}
\maketitle

Quantum mechanical systems can undergo zero-temperature phase transitions upon variation of a non-thermal control parameter, where the order is destroyed solely by quantum fluctuations.\cite{Subir} In this Letter, we re-explore second-order quantum phase transitions in impurity models which involve a spin coupled to a dissipative environment and which display both a localized and a delocalized phase for the spin.\cite{Matthias} To better characterize those quantum phase transitions, we are prompted to examine the entropy of entanglement shared between the spin and its environment.\cite{Bennett,Amico} As a concrete example, we employ the subohmic spin-boson model\cite{Matthias} describing a two-level system with environmental dissipation (stemming from a lossy $RLC$ transmission line\cite{Matthias2} or from $1/f$ noise\cite{noise}) and which allows a direct measurement of the entropy of entanglement. We demonstrate that the ground state is strongly entangled at the quantum phase transition. The analysis of the enhancement of entanglement  near a quantum critical point is of great current interest.\cite{Amico,Osborne,rest}

More precisely, we show that those second-order impurity quantum phase transitions are always accompanied by a maximum (cusp) in the entropy of entanglement and that the latter exhibits universal scalings. The entanglement entropy will also allow us to establish important connections between impurity entanglement, quantum decoherence (or strong reduction of the quantum superposition of the two spin states) when approaching the  phase transition from the delocalized phase, and rapid disentanglement in the localized or classical phase for the spin (the spin is rapidly frozen in one classical state due to dissipation). Our phase diagram is shown in Fig. 1. 

\begin{figure}[ht]
\includegraphics[width=2.65in,height=1.9in]{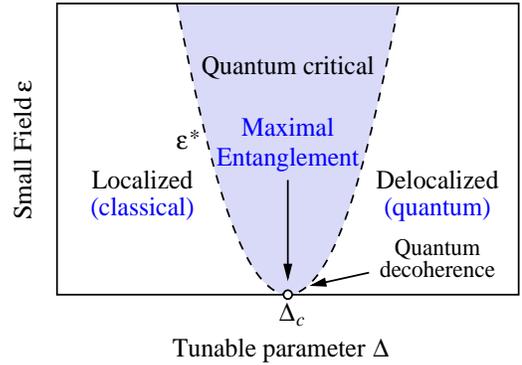}
\caption{\label{crossover} (color online) Unification of entanglement and criticality at a second-order impurity quantum phase transition.}
\end{figure}

The simplest example of a quantum impurity coupling to a bath is the paradigmatic spin-boson model: \cite{Leggett}
\begin{equation}
H_{SB}=-\frac{\Delta}{2}\sigma_x+\frac{\epsilon}{2}\sigma_z+H_{osc}+\frac{1}{2}\sigma_z \sum_n \lambda_n(a_n+a_n^{\dagger}),
\label{Hsb}
\end{equation}
where $\sigma_x$ and $\sigma_z$ are Pauli matrices and $\Delta$ is the tunneling amplitude between the states with $\sigma_z=\pm 1$.  In the following, we will assume that $\Delta>0$ such that $\langle \sigma_x\rangle >0$. Here, $H_{osc}$ is the Hamiltonian of an infinite number of harmonic oscillators with frequencies $\{ \omega_n \}$, which couple to the spin degree of freedom via the coupling constants $\{ \lambda_n \}$.  The heat bath is characterized by its spectral function,
${\cal J}(\omega) \equiv \pi \sum_n \lambda_n^2 \delta (\omega_n-\omega) =2\pi \alpha \omega_c^{1-s} \omega^s$, where $\omega_c$ represents a cutoff energy and the dimensionless parameter $\alpha$ measures the strength of the dissipation. 

The essential physics contained in this model is the competition between the amplitude for tunneling between the two spin states (leading to a ``delocalized'' phase) and the effect of the bath which tends to
``localize'' the system in one or other of the spin states. 

The special value $s=1$ represents the case of ohmic dissipation, where the analogy with the Kondo model applies and a quantum Kosterlitz-Thouless (KT) transition separates the delocalized phase at small $\alpha$ and the localized phase at large $\alpha$.\cite{Leggett} The phase transition occurs at $\alpha_c=1+{\cal O}(\Delta/\omega_c)$.\cite{Cha} The delocalized region corresponds to the antiferromagnetic Kondo model with a Fermi-liquid ground state while the localized region corresponds to the ferromagnetic Kondo model where the spin is frozen in time. The spin magnetization $\langle \sigma_z\rangle$ jumps by a non-universal amount $-1+{\cal O}(\Delta/\omega_c)$ at the KT transition (for $\epsilon=0^+$).\cite{Anderson} 
The entropy of entanglement also contains valuable information; it shows a plateau at maximal entanglement for $\alpha\geq 1/2$ in the delocalized phase and it drops (roughly) to zero at the KT transition.\cite{AngelaKaryn,Costi,Angela} Note that this plateau exemplifies the intimate connection between maximal entanglement and quantum decoherence since $\alpha=1/2$ marks the dynamical crossover from damped oscillatory to overdamped behavior.\cite{Leggett}

Below, we rather focus on the subohmic spin-boson model $(0<s<1)$. A quantum critical point $\Delta_c(\alpha)$ still separates the localized $(\Delta<\Delta_c)$ and the delocalized phase $(\Delta>\Delta_c)$ 
where $\Delta_c\rightarrow 0$ when $\alpha\rightarrow 0$. A second-order phase transition has been well-established for all $0<s<1$ using Wilson's Numerical Renormalization Group (NRG) method\cite{Bulla,Meirong} and through an analogy to classical spin chains\cite{Dyson}  (nevertheless, when interactions become too long-ranged in time, {\it i.e.}, for $s<1/2$, the quantum-classical mapping fails\cite{Matthias}). The longitudinal spin magnetization $\langle \sigma_z \rangle$ is generally used to characterize the phases and phase transitions.\cite{Qimiao2} On the other hand, the quantum critical point is an ``interacting'' fixed point for all $0<s<1$ which motivates us to examine the entanglement between the spin and its environment thoroughly. 
When a bipartite quantum system AB is in a pure state $|\psi\rangle$, the entanglement  between subsystems A and B is unambiguously described by the {\it von Neumann} entropy or entanglement entropy $E$ which is calculated from the reduced density matrix $\rho_A$ or 
$\rho_B$, $\rho_{A(B)}=\mathrm{Tr}_{B(A)} | \psi\rangle\langle\psi |$:\cite{Bennett}
\begin{equation}
 E=-\mathrm{Tr} \left(\rho_A \log_2 \rho_A\right)=-\mathrm{Tr} \left(\rho_B \log_2 \rho_B\right).
 \end{equation}
 When either subsystem A or B is a spin-$\frac{1}{2}$ system, the entropy of entanglement $E$ can be rewritten as,\cite{Amico,AngelaKaryn,Costi,Angela}
 $E=-p_+ \log_2 p_+ - p_- \log_2 p_- $, 
 where $p_{\pm}=\left( 1\pm \sqrt{\langle \sigma_x \rangle^2 + \langle \sigma_z \rangle^2}\right)/2$; note that $\langle \sigma_y \rangle =0$ because the Hamiltonian $H_{SB}$ is invariant under $\sigma_y \to -\sigma_y$. 

Since exact analytical methods (such as Bethe Ansatz) are restricted to the ohmic case,\cite{AngelaKaryn,Markus} we employ the bosonic NRG to compute the entanglement entropy between the spin and its environment. We follow the same procedure as in our Ref. \onlinecite{Meirong}. In order to ensure the convergence of the results in the localized phase, we have used  the  NRG parameters $\Lambda=2$ (logarithmic discretization), $N_s=150$ (lowest energy levels kept), and $N_b=8$ (boson states; except the 0th site for which $N_{b0}=500$). The results converge for $N\approx 30$ sites (and we keep until 40 sites). In NRG calculations, we fix $\omega_c=1$ and $\Delta$ is normalized to $\omega_c$.
 \begin{figure}[ht]
\includegraphics[width=3in,height=2in]{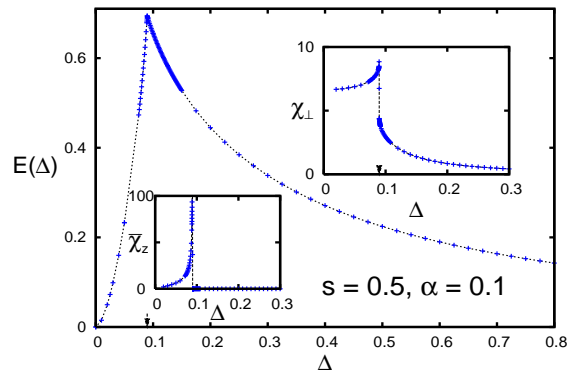}
\caption{\label{crossover} (color online) NRG results for the entanglement at $T=\epsilon=0$ in the subohmic case $s=0.5$, which can be realized through a qubit coupled to an $RLC$ transmission line; the arrow marks the position of the quantum phase transition. The relevant susceptibilies $\bar{\chi}_z$ and $\chi_{\perp}$ are presented in insets.}
\end{figure}
We will also apply the following scaling ansatz for the impurity part of the free energy,
\begin{equation}
\label{free}
F=Tf(|\Delta-\Delta_c|/T^{1/\nu},\epsilon T^{-b}),
\end{equation}
to relate the critical exponents associated with $E$ to other critical
exponents such as the correlation length exponent $\nu$. Even though the temperature $T$ is introduced for the scaling analysis the entanglement entropy $E$ (which is defined for a pure state) is evaluated at zero temperature. The crossover from the quantum critical regime to one or other of the stable regimes, defines the energy scale $\epsilon^*$ in Fig. 1 that vanishes at $\Delta_c$ as $\epsilon^*\propto |\Delta_c-\Delta|^{b\nu}$, and for the spin-boson model we obtain $b=(1+s)/2$. The ansatz (\ref{free}) is well justified when the fixed point is interacting;\cite{Qimiao2} for a Gaussian fixed point the scaling function would also depend upon dangerously irrelevant variables.

Note that the entanglement entropy $E$ is different from the impurity entropy $S_{imp}$ which is rather evaluated as entropy of the system with impurity minus entropy of the bath alone. The delocalized phase obeys $S_{imp}(T\rightarrow 0)=0$ since the ground state is non-degenerate while the localized phase yields $S_{imp}(T\rightarrow 0)=\ln 2$. On the other hand, $E$ goes rapidly to zero in the localized phase  (product state) and is finite in the delocalized phase (entangled state); however, $E\rightarrow 0$ when $\Delta\gg \omega_c$. Our NRG data of Figs. 2 and 3 show that $E$ exhibits a cusp-like behavior (maximum) at the phase transition. This is distinguishable from the ohmic case where $|\langle \sigma_z\rangle| = 1 -{\cal O}(\Delta/\omega_c)$ at the KT transition which irrefutably results in $E\rightarrow 0$.\cite{Costi,Angela}

In fact, there is a simple way to conceive that the maximum of $E$ coincides with the quantum phase
transition for the subohmic situation. In the delocalized phase, since the longitudinal magnetization 
$\langle \sigma_z\rangle=0$ at $\epsilon=0$:
\begin{equation}
\frac{\partial E}{\partial \Delta} = -\frac{\chi_{\perp}}{2\ln 2} \ln\left[\frac{1+\langle \sigma_x\rangle}{1-\langle \sigma_x\rangle}\right].
\end{equation}
This expression is valid at $\Delta_c$.  Since  $\chi_{\perp}=\partial \langle \sigma_x\rangle/\partial \Delta$ and  $\langle \sigma_x\rangle$ are positive quantities, $\partial E/\partial \Delta <0$ in the delocalized phase. In the localized phase, $E$ is controlled by the finite longitudinal magnetization and by the susceptibility 
$\bar{\chi}_z=-\partial |\langle \sigma_z\rangle|/\partial \Delta>0$ (see inset in Fig. 2):\cite{footnote1}
\begin{equation}
\frac{\partial E}{\partial \Delta} \approx \frac{\bar{\chi}_{z} |\langle \sigma_z\rangle|}{2\ln 2 \langle\sigma_x\rangle} \ln\left[\frac{1+\langle \sigma_x\rangle}{1-\langle \sigma_x\rangle}\right].
\end{equation}
\begin{figure}[ht]
\includegraphics[width=3.2in,height=2.3in]{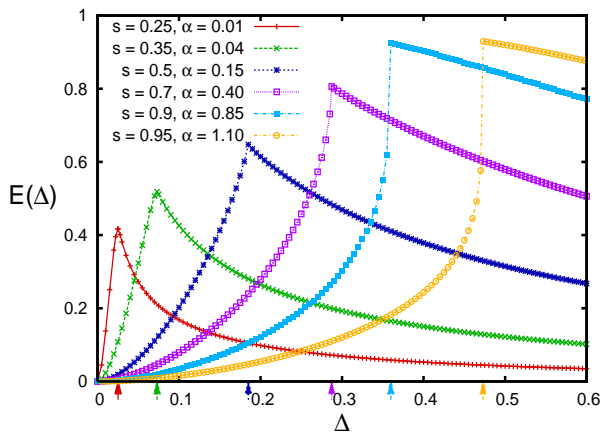}
\caption{\label{crossover} (color online) The entropy of entanglement $E$, obtained from NRG, displays a cusp-like behavior for all $0<s<1$. For $s\rightarrow 1$, $E$ becomes rapidly suppressed at $\Delta_c^-$ which is a reminiscence of the ohmic case (KT transition).}
\end{figure}

\hskip -0.3cm Thus, $\partial E/\partial \Delta >0$ in the localized phase. Eqs. (6) and (7) imply that the entanglement entropy is maximum at the phase transition. Zero-temperature impurity critical points can show a residual ``fractional'' entanglement entropy (which depends on the dissipation strength $\alpha$).

Let us emphasize at this point that the case $s=1/2$ is of particular interest since it can be realized  through a charge qubit (dot) subject to the electromagnetic noise of an $RLC$ transmission line;\cite{Matthias2} the mapping onto the spin-boson model with $s=1/2$ assumes frequency $\omega\ll R/L$, whereas an $LC$ transmission line would mimic the ohmic case.\cite{Markus,Karyn2,Karyn3} The entanglement entropy can be accessed through charge and persistent current measurements, corresponding to $\langle \sigma_z\rangle$ and $\langle \sigma_x\rangle$ respectively.\cite{Markus} Moreover, $\Delta$ represents the tunneling amplitude between the dot and the lead or the Josephson energy of the junction and thus is a tunable parameter. A gate 
can be used to control $\epsilon$. The noisy charge qubit is an ideal Nano-candidate to demonstrate the existence of entanglement.\cite{AngelaKaryn,Jordan}

Now, we seek to demonstrate that $E$ exhibits universal scaling even though the entanglement is two-sided, so that two numbers are necessary to specify $E$ (one for each of the two ways of approaching $\Delta_c$). Near $\Delta_c$, the transverse spin susceptibility $\chi_{\perp}=2\partial^2 F/(\partial \Delta)^2$ obeys 
\begin{equation}
\label{chi}
\chi_{\perp}(\Delta)= \chi_{\perp}(\Delta_c) +c_{+/-}|\Delta-\Delta_c|^{\zeta},
\end{equation}
and from the free energy defined in Eq. (\ref{free}),  $\zeta=\nu-2$. For the subohmic spin-boson model,  one finds $\nu\geq 2$ for all $0<s<1$, ensuring that $\chi_{\perp}$ does not diverge at the transition. 
Taking into account that $\langle \sigma_x\rangle$ is continuous at the transition, Eq. (\ref{chi}) thus implies that $E$ always rises linearly for $\Delta\rightarrow \Delta_c^+$
($\Delta_c^{\pm}$ means that we approach the quantum critical point from the delocalized (localized) region), as confirmed by Fig. 3.  Note that the coefficients $c_+$ and $c_-$ can be different in the delocalized and in the localized phase and this explains for example the jump occurring for $s=1/2$ where $\nu=2$ (inset in Fig. 2). 
\begin{figure}[ht]
\includegraphics[width=3.2in,height=2.3in]{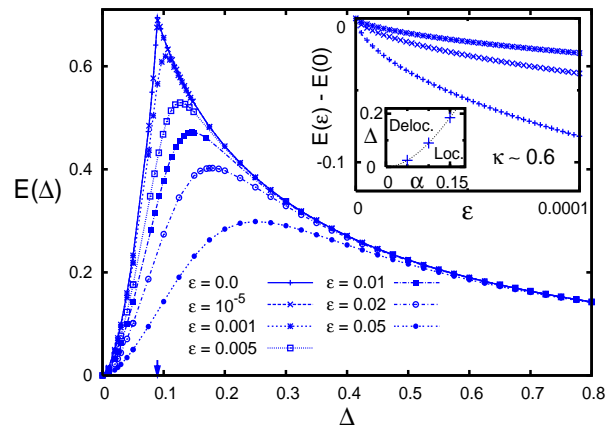}
\caption{\label{crossover} (color online) $E(\Delta)$ at $s=0.5$ and $\alpha=0.1$ for several values of $\epsilon$. Inset: universal scaling in the critical region for three different values of $\Delta_c$ (from NRG  we obtain $\kappa\sim 0.6$ and the exact exponent is $\kappa=2/3=0.66...$).}
\end{figure}
Through the NRG, we also check that $c_+<0$, emphasizing that in the delocalized phase $\chi_{\perp}$ substantially increases at $\Delta_c^+$, and that $\chi_{\perp}$ shows a clear singularity for $2\leq \nu<3$ or $1/3<s\leq 0.94$. For all $0<s<1$, this strongly underlines the duality between the enhancement of entanglement and the strong reduction of the two-spin state quantum superposition near the  phase transition.

In the localized phase, we obtain the scaling behavior:
\begin{equation}
\bar{\chi}_z(\Delta) \propto  |\Delta-\Delta_c|^{-1+\nu(1-s)/2} + a;
\end{equation} 
here $a\neq 0$ when $\nu(1-s)/2>1$, and we identify $a=\bar{\chi}_z(\Delta_c^-)$. For $\Delta>\Delta_c$, $\bar{\chi}_z=0$. The correlation length exponent $\nu$ can be obtained analytically for $s\rightarrow 0$ and $s\rightarrow 1$ through Renormalizations Group expansions. The analogy with classical spin chains for $s\rightarrow 1$ leads to\cite{Kosterlitz} $1/\nu=\sqrt{2(1-s)}$ whereas at small $s$, one finds\cite{Matthias} $1/\nu=s$. Here, $\bar{\chi}_z$ diverges at $\Delta_c^-$ for $s>1/3$ and $\beta=\nu(1-s)/2$ is the critical exponent associated to 
$\langle \sigma_z\rangle$. For $\Delta<\Delta_c$ and $s>1/3$, from Eqs. (7) and (9), we find the relation\cite{footnote3} \begin{equation}
E(\Delta_c) - E(\Delta) \propto |\Delta-\Delta_c|^{\nu(1-s)}.
\end{equation}
The decay of the von Neumann entropy $E$ in the localized phase is faster than linear for all $s>1/2$ (Fig. 3) and the behavior becomes strictly linear at $s=1/2$, as shown in Fig. 2. It is certainly relevant to notice the parallel between impurity entanglement in a dissipative environment and single-site entanglement in quantum critical spin chains such as the anisotropic XY chain.\cite{Osborne} 
Now, we shall discuss the scaling of $E$ with the longitudinal field. 

 Integrating out the boson degrees of freedom induces a long-range interaction in time which results in the following term in the action,\cite{Matthias} ${\cal S}_{int} = \int d\tau d\tau' \sigma_z(\tau) g(\tau-\tau') \sigma_z(\tau')$,
where $g(\tau)\propto 1/\tau^{1+s}$ at long times. Assuming that the dynamics of $\sigma_z$
at the critical point is essentially determined by $S_{int}$ and by the local field $\epsilon$ we then derive
$\langle \sigma_z\rangle (\epsilon,\Delta_c) \propto |\epsilon|^{1/\delta}$,
\begin{figure}[ht]
\includegraphics[width=3in,height=2.2in]{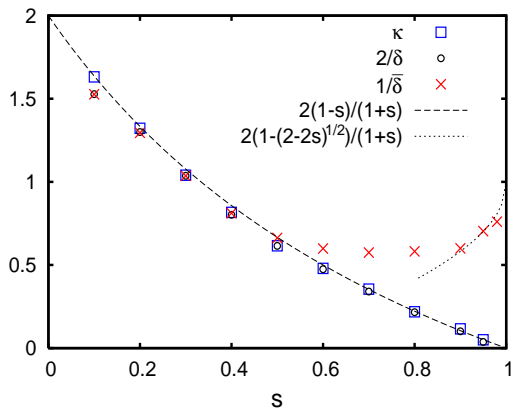}
\caption{\label{crossover} (color online) Exponent $\kappa$ versus $s$ from NRG. For all $0<s<1$, $\kappa=2/\delta$ and for $0<s\leq 1/2$, 
$\kappa=2/\delta=1/\bar{\delta}$.}
\end{figure}
with the exponent $\delta=(1+s)/(1-s)$. This is consistent with our NRG results which predict that the
local susceptibility $\chi_z=\partial |\langle \sigma_z\rangle|/\partial\epsilon$ at the quantum critical point diverges as $T^{-s}$. In the small
$s$ limit, this leads to $1/\delta \approx 1-2s+{\cal O}(s^2)$; this can be recovered by resorting to a small $s$ expansion.\cite{Matthias} Using $\langle \sigma_x\rangle=-2\partial F/\partial \Delta$,  we obtain
$\langle \sigma_x \rangle (\epsilon,\Delta_c) - \langle \sigma_x \rangle (\Delta_c)
\propto -|\epsilon|^{1/\bar{\delta}}$, with $1/\bar{\delta}=\frac{2}{1+s}(1-1/\nu)$.
To maintain consistency with our notations,  we identify $\langle \sigma_x \rangle (\epsilon=0,\Delta_c)=\langle \sigma_x \rangle (\Delta_c)$, and similarly for other quantities (observables).

When $s$ is close to one, the critical point is defined by $\alpha=1$ and
$\Delta_c^2=1-s$. This ensures that $\langle \sigma_x\rangle$ is small at the transition and evolves slowly with $\epsilon$; at $s=1$, around the phase transition, one gets the exact expression $\langle \sigma_x\rangle(\epsilon=0) =\Delta/[\omega_c(2\alpha-1)]$.\cite{AngelaKaryn} Thus, when $s$ is close to one, the dependence of $E$ on $\epsilon$ mainly stems from $\langle \sigma_z\rangle$:
\begin{equation}
E(\epsilon,\Delta_c)-E(\Delta_c) \propto -|\epsilon|^{\kappa},
\end{equation}
and $\kappa=2/\delta=2(1-s)/(1+s)$. In fact, since $1/\bar{\delta}(s)\geq 2/\delta(s)$, this scaling relation remains valid for all $0<s<1$, as shown in Figs. 4 (inset) and 5; one can always expand $p_{\pm}(\epsilon)=p_{\pm}(\epsilon=0) \pm m\epsilon^{2/\delta}$ at small $\epsilon$ and $m>0$ to satisfy $\partial_{\epsilon } E(\epsilon)<0$ (the field $\epsilon$ favors a product state). On the other hand, since at small $s$ the critical exponent $\nu$ obeys $\nu=1/s$, one also gets $1/\bar{\delta}=2/\delta=\kappa$, which is well verified through the NRG for $0<s\leq 1/2$ (Fig. 5).  

For $\Delta>\Delta_c$, the NRG results predict $\langle \sigma_z\rangle(\epsilon,\Delta) \propto \epsilon$ and $\langle \sigma_x\rangle(\epsilon,\Delta) - \langle \sigma_x\rangle(\Delta) \propto -\epsilon^2$. Thus, $E$ decreases as $\epsilon^2$ similar to the ohmic case.\cite{AngelaKaryn} Since $2/\delta<2$ for all $0<s<1$, this implies that for a 
given $\epsilon\neq 0$, the maximum of entanglement occurs at the value of $\Delta>\Delta_c$ which
lies in the crossover between the delocalized and the quantum critical regime (Figs. 1 and 4). The delocalized phase is quite robust to the application of a  field $\epsilon$. For $\Delta<\Delta_c$, in contrast we find a linear decrease of $E$ with $\epsilon$.

In conclusion, we have shown that the entanglement between a spin and its (bosonic) environment is always enhanced at a second-order quantum phase transition. The concept of entanglement entropy allows us to establish important connections between impurity entanglement, strong reduction of the quantum superposition of the two spin states when approaching the phase transition from the delocalized phase, rapid disentanglement in the localized phase, and criticality. Our theoretical results can be tested experimentally through a charge qubit coupled to a lossy $RLC$ transmission line. These results may also be relevant for heavy fermion systems which might develop a similar ``local'' criticality.\cite{Qimiao} 

We thank A. Kitaev for useful discussions. P.D.-B. and K.L.H. acknowledge financial support from NSERC.


\begin{thebibliography}{10}
\bibitem{footnote1}
 Here, we neglect the less relevant contribution from $\chi_{\perp}$.
\end{thebibliography}
\end{document}